\documentclass[aps,prl,showpacs,preprintnumbers,twocolumn,superscriptaddress]{revtex4-2}
\usepackage{amsmath,amssymb}
\usepackage{bm}
\usepackage{tipa}
\usepackage{upgreek}
\usepackage{comment}
\usepackage{mathrsfs}
\usepackage{graphicx}
\usepackage{braket}
\usepackage{enumitem}
\usepackage{mathbbol}
\usepackage{booktabs}
\usepackage{gensymb}
\usepackage[normalem]{ulem}
\usepackage{color}
\usepackage[colorlinks,bookmarks=true,citecolor=blue,linkcolor=red,urlcolor=blue]{hyperref}
\usepackage{hyperref}
\renewcommand{\vec}[1]{\mathbf{#1}}
\usepackage{pifont}

\newcommand{\bk}{\bm{k}}

\newcommand{\bp}{\bm{p}}
\newcommand{\bq}{\bm{q}}

\begin{document}

	\title{Superconductivity from repulsive interactions in Bernal-stacked bilayer graphene}
	\author{Glenn Wagner}
	\affiliation{Department of Physics, University of Zurich, Winterthurerstrasse 190, 8057 Zurich, Switzerland}
	\author{Yves H. Kwan}
	\affiliation{Rudolf Peierls Centre for Theoretical Physics, Parks Road, Oxford, OX1 3PU, UK}
	\affiliation{Princeton Center for Theoretical Science, Princeton University, Princeton NJ 08544, USA}
	\author{Nick Bultinck}
	\affiliation{Rudolf Peierls Centre for Theoretical Physics, Parks Road, Oxford, OX1 3PU, UK}
	\affiliation{Department of Physics, Ghent University, Krijgslaan 281, 9000 Gent, Belgium}
	\author{Steven H. Simon}
	\affiliation{Rudolf Peierls Centre for Theoretical Physics, Parks Road, Oxford, OX1 3PU, UK}
	\author{S.A. Parameswaran}
	\affiliation{Rudolf Peierls Centre for Theoretical Physics, Parks Road, Oxford, OX1 3PU, UK}

	\begin{abstract}
    A striking series of experiments have observed superconductivity in Bernal-stacked bilayer graphene (BBG) 
    when the energy bands are flattened by applying an electrical displacement field. Intriguingly, superconductivity manifests only at non-zero magnetic fields, or when spin-orbit coupling is induced in BBG by coupling to a substrate. We present detailed functional renormalization group and random-phase approximation calculations that provide a unified explanation for the superconducting mechanism in both cases. Both calculations yield a purely electronic $p$-wave instability of the Kohn-Luttinger (KL) type. The latter can be enhanced either by magnetic fields or Ising spin-orbit coupling, naturally explaining the behaviour seen in experiments.

	\end{abstract}
	
 	\maketitle

\textit{Introduction}.---The explosion of interest in magic-angle twisted bilayer graphene (TBG), sparked by experimental observations of gate-tunable superconductivity (SC)~\cite{Cao2018,Cao2018b,Yankowitz_2019,Lu2019} and correlated insulating behaviour~\cite{Park_2021,Cao2018,Yankowitz_2019,Cao_2021,liu2021tuning,Sharpe_2019,Serlin900,Lu2019,Stepanov_2020,Wu_2021,Zondiner_2020,pierce2021unconventional,polshyn2019large,uri2020mapping,saito2020independent,Das_2021,saito2021isospin,rozen2021entropic,stepanov2020competing}, has stimulated broader investigations of correlated electron physics in two-dimensional materials with narrow energy bands. Recently, several  systems of moir\'eless graphene multilayers have been shown to host correlation effects reminiscent of their more complex cousins. Most notably, the application of an electrical displacement field  to Bernal-stacked bilayer graphene (BBG) and rhombohedral trilayer graphene (RTG) flattens the bands near neutrality and  gate-tunable SC has been observed in both BBG  \cite{SC_BLG,BBG_SOC_Exp} and RTG \cite{RTG_SC} in such a setting. In the phase diagram of both systems, the superconductor is proximate to a cascade of symmetry-breaking transitions, again a feature familiar from TBG.

 Experimentally, RTG exhibits a cascade of symmetry-breaking transitions and SC with a critical temperature $T_c=106$~mK.  As in TBG, the symmetry-breaking transitions can be explained within a Hartree-Fock mean-field approximation~\cite{HuangCascadesRTG}, while candidate theories of SC in RTG range from purely electronic mechanisms~\cite{Ghazaryan2021,Ghazaryan2022,Chatterjee2021,You2022,SzaboRTG,hotspotRG,qin2022FRG,Cea2022,pantaleon2022related} to acoustic-phonon-mediated attraction~\cite{Chou1}.

Similar to TBG and RTG, quantum oscillation measurements in BBG show a variety of isospin symmetry breaking transitions \cite{delaBarrera2022,Seiler_2022}, that can be understood in terms of Stoner ferromagnetism \cite{Dong2021,Szabo}. In contrast to those systems, however, BBG becomes superconducting  only  in the presence of either an in-plane magnetic field \cite{SC_BLG} or  spin-orbit coupling (SOC) induced by placing the BBG on top of a layer of WSe$_2$ \cite{BBG_SOC_Exp}, with  transition temperatures of $T_c=26$~mK and $T_c=260$~mK respectively. 
Although both phonon-mediated \cite{Chou1,ChouBBG,chou2022enhanced} and purely electronic SC mechanisms have been proposed for BBG \cite{Szabo,BBG_KL,Dong2022,cea2022intervalley,pantaleon2022related}, the requirement of a magnetic field or SOC to trigger SC is a new ingredient, absent in either TBG or RTG, that could help pinpoint  
the nature of the SC instability. A possible explanation based on fluctuating superconductivity was proposed in Ref.~\cite{Curtis}, which remained agnostic as to the origin of the pairing ``glue''.

A Kohn-Luttinger mechanism provides a potential pathway for superconductivity from purely repulsive interactions,  by generating an effective interaction in higher angular momentum channels via overscreening of the Coulomb interaction \cite{KL}, yet its viability and the precise features of the resulting SC will depend sensitively on details of the underlying Fermi liquid parent state. To address this question quantitatively,  we study Kohn-Luttinger type SC in BBG via the random-phase approximation (RPA) and functional renormalization group (FRG) calculations. The results from both approaches can be described within a simplified three-pocket model that captures the essential features of the BBG Fermi surface. By incorporating the effects of applied field and SOC, we show that an all-electronic superconducting mechanism  provides a unified explanation for both classes of experiment.

\begin{figure}[h]
    \centering
    \includegraphics[width=\columnwidth]{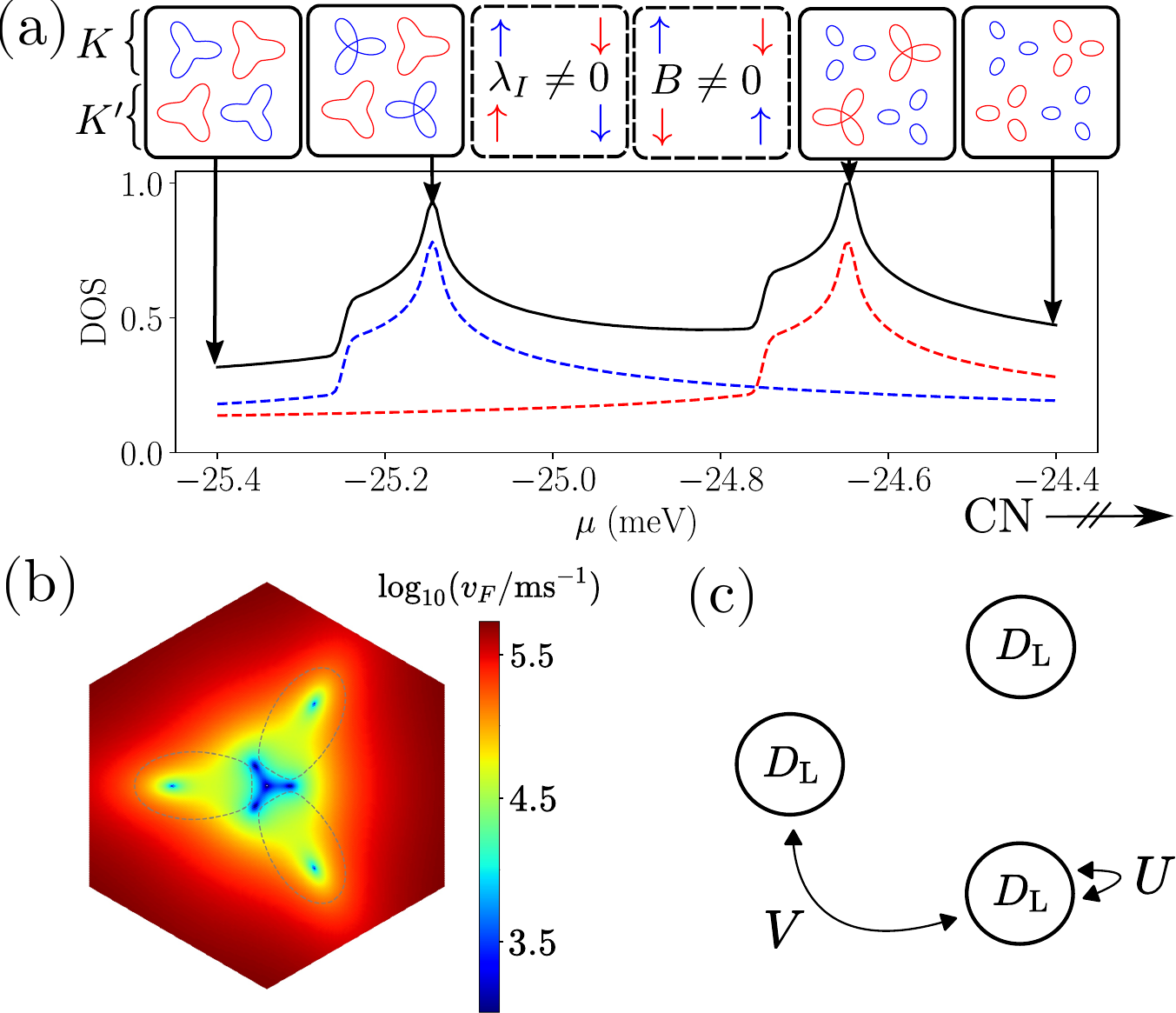}
    \caption{(a) Density of states (DOS) as a function of the chemical potential $\mu$  in the presence of Ising SOC $\lambda_I=0.5$meV or equivalently an in-plane field of $B=4.3$T. Red and blue dashed lines are  contributions of  distinct symmetry-related Fermi surfaces (FSs) of the four spin/valley species shown in the solid boxes to the total DOS (black solid line). The $B$-field and  Ising SOC lead to distinct assignments of spin labels to the FSs (dashed boxes). The vHS closer to charge neutrality (CN, $\mu=0$) corresponds to the majority isospin species, whereas the one further from CN  corresponds to the the minority isospin species. Strong SC is only observed near the former. We include thermal broadening of $T=0.1$K. (b) Fermi velocity in the Brillouin zone close to the $K$-point with dashed FS for $B=0$ and $\mu=-24.92$meV. (c) Pocket model derived from three pockets with inter-pocket ($V$) and intra-pocket ($U$) interactions.}
    \label{fig:DOS}
\end{figure}

\textit{Hamiltonian}.---
We begin with a low-energy four-band model for electrons in BBG~\cite{McCann2006,Jung2014}. In the basis $\{1A,1B,2A,2B\}$ (where the number indicates the layer and $A/B$ label distinct sublattices in a single  layer) the low-energy Hamiltonian is 
\begin{equation}
\label{eq:Ham}
H=\left(\begin{array}{cccc}
\frac{D}{2} & v_0 \pi^{\dagger} & -v_{4} \pi^{\dagger} & -v_{3} \pi \\
v_0 \pi & \Delta^{\prime}+\frac{D}{2}& t_{1} & -v_{4} \pi^{\dagger} \\
-v_{4} \pi & t_{1} & \Delta^{\prime}-\frac{D}{2} & v_0 \pi^{\dagger} \\
-v_{3} \pi^{\dagger} & -v_{4} \pi & v_0 \pi & -\frac{D}{2}
\end{array}\right),
\end{equation}
where $\pi=\hbar(\tau_z k_x+ik_y)$ and $s_i$ and $\tau_i$ denote Pauli matrices associated with spin and valley respectively. The displacement field is chosen to be $D=50\,$meV and $v_i=t_{i} \sqrt{3} a / 2 \hbar$ with $a=0.246\,$nm the lattice constant of graphene. We use the tight-binding parameters~\cite{Jung2014}, $t_0=2.61\,$eV, $t_1=0.361\,$eV, $t_3=0.283\,$eV, $t_4=0.138\,$eV and $\Delta'=0.015\,$eV. $t_3$ controls the trigonal warping. We add the gate-screened Coulomb interaction $V^0(q)=\frac{e^2}{2\epsilon_0\epsilon_r q}\tanh{qd_\textrm{sc}}$ with screening length $d_\textrm{sc}$ and relative permittivity $\epsilon_r$. We choose a UV cutoff to more finely resolve details 
of the Fermi surface \footnote{In practice, we pick a UV cutoff of $0.025\times (\pi/a)$ where $a$ is the lattice constant of graphene.} and neglect the Bloch form factors for simplicity (we anticipate that these only affect the physics quantitatively). We also neglect the weak intervalley exchange scattering, such that our model has separate spin rotation symmetry in each valley, i.e.~$SU(2)_K\times SU(2)_{K'}$. In this limit, the magnetic field Zeeman term $\mu_BB_Zs_z$ and the Ising SOC term $\frac{\lambda_I}{2}s_z\tau_z$ are equivalent (up to a flavour rotation) and we can treat the SOC as an effective Zeeman field $B_\textrm{SOC}=\lambda_I/(2\mu_B)$. In the presence of a Zeeman field the density of states of  majority and minority isospin species exhibit van Hove singularities (vHS)
at different chemical potentials (see Fig.~\ref{fig:DOS}). 

\begin{figure*}
    \centering
    \includegraphics[width=0.9\textwidth]{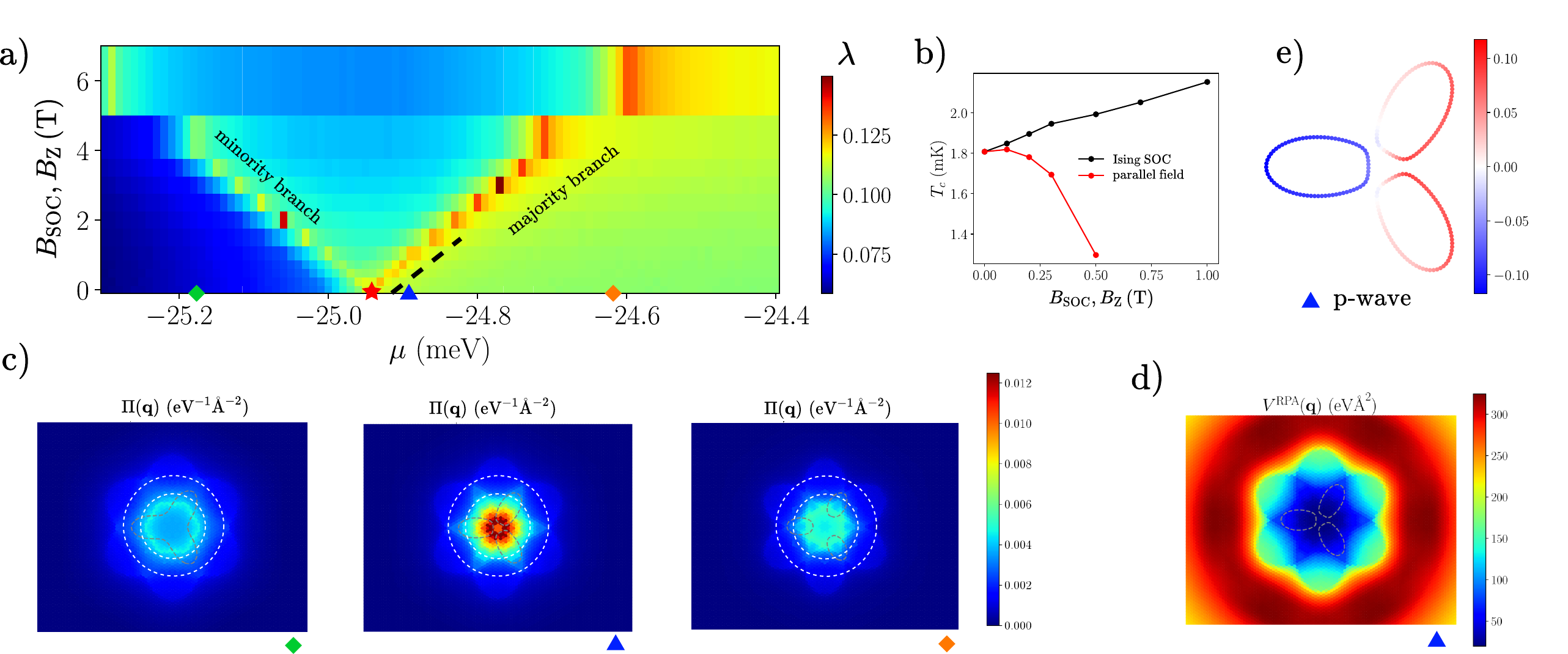}
    \caption{\textbf{Kohn-Luttinger superconductivity in the random phase approximation.} \textbf{a)} Maximum gap matrix eigenvalue $\lambda$, related to the superconducting critical temperature $T_c\sim E_0e^{-\frac{1}{\lambda}}$, as a function of Zeeman field $B$ and chemical potential $\mu$. \textbf{b)} Comparison of the effects of orbital coupling due to an in-plane magnetic field on $T_c$. Horizontal axis corresponds to the dotted black line in a). Energy cutoff of the gap equation set as $E_0=10\,\text{meV}$. \textbf{c)} Static polarization function in the absence of a (generalized) Zeeman field at three dopings indicated by the corresponding symbols in a). The annular region between the dashed white lines indicates the momentum range that contributes to $\Pi(q_P)$, where $q_P$ is the inter-pocket momentum. \textbf{d)} RPA-screened interaction at the chemical potential indicated with a blue triangle in a). Bare interaction is of the dual gate-screened form with relative permittivity $\epsilon_r=5$ and screening distance $d_{\text{sc}}=38\,\text{nm}$ and we set $\Delta'=0$meV. Fermi surface in valley $\tau=+$ is shown with grey dashed contours. \textbf{e)} Representative gap function corresponding to $p$-wave superconductivity. The solution is two-fold degenerate (we show the $p_x$ solution).}
    \label{fig:RPA_main}
\end{figure*}

\textit{Pocket model.---} 
A key outcome of our detailed numerical simulations is that  the key features of SC in BBG can be captured within 
a simplified three-pocket model, that we now describe to orient our discussion (and justify {\it a posteriori}, via our RPA/FRG calculations). The distinguishing feature of the Fermi surface shown in Fig.~\ref{fig:DOS}c is the presence of three pockets related by $C_3$ symmetry each with density of states $D_L$. The key physics is then controlled by couplings $U,V$ that represent the intra-pocket and inter-pocket interactions respectively. The gap equation takes the form $\sum_{\mathbf{k}'}M_{\mathbf{k},\mathbf{k}'}\Delta_{\mathbf{k}'}=\lambda\Delta_{\mathbf{k}}$, where $\mathbf{k}$ runs over $N_p$ momenta lying on the Fermi surface. Assuming for now that the interactions are the same for all the momenta within one pocket, the gap matrix simplifies to a $3\times3$ matrix
\begin{equation}
    M=-\begin{pmatrix} U & V & V\\V&U&V\\V&V&U
    \label{eq:M}
    \end{pmatrix},
\end{equation}
where we have neglected dimensional and normalization factors. A positive eigenvalue $\lambda$ indicates a superconducting instability with $T_c\sim E_0e^{-\frac{1}{\lambda}}$, where $E_0$ is a UV cutoff. $M$ has leading eigenvectors $\Delta\sim(1, e^{\pm\frac{2\pi i}{3}}, e^{\pm\frac{4\pi i}{3}})^T$ corresponding to a degenerate $p$-wave solution with eigenvalue $\lambda_p=(V-U)$.  The bare Coulomb interaction is monotonically decreasing as a function of momentum and therefore at the bare level $V^0<U^0$ and there is no superconductivity. However, due to screening we can have $V>U$ such that we obtain superconductivity. Within the RPA, screening leads to
\begin{equation}\label{eqn:lambda_p}
    \lambda_p=\frac{V^0}{1+\Pi(q_{P})V^0}-\frac{U^0}{1+\Pi(0)U^0},
\end{equation}
where $q_{P}$ is the inter-pocket distance, $\Pi(0)$ reflects the total DOS, and $\Pi(q_{P})$ is roughly the average of the polarization function within the annular region in Fig.~\ref{fig:RPA_main}d. Since  $\Pi(0)>\Pi(q_{P})$, we can obtain $\lambda_p>0$ after the screening.

\textit{RPA.---} To demonstrate that a Kohn-Luttinger-like mechanism can lead to superconductivity in BBG, we perform an initial 
RPA analysis. An RPA analysis on BBG was already performed in~\cite{BBG_KL}, where nodal $s$-wave and $p$-wave solutions are obtained and SOC is seen to enhance the critical temperature.
In our calculation we use a patching scheme that involves dividing the Fermi surface into finite segments, with the dispersion in the direction perpendicular to the Fermi surface  treated in the linear approximation and integrated (with UV cutoff $E_0$) to obtain the logarithmic Cooper divergence. The largest positive eigenvalue $\lambda$ of the the symmetrized gap matrix corresponds to a superconducting solution with $T_c\sim E_0e^{-\frac{1}{\lambda}}$. The details of the RPA calculation are provided in \cite{Supplement}.

Fig.~\ref{fig:RPA_main}a charts the maximum eigenvalue $\lambda$ as a function of chemical potential and applied Zeeman field. Focusing first on $B=0$, we find that superconductivity exists for all values of $\mu$ shown, despite purely repulsive electronic interactions. The maximum $T_c$ is attained around the vHS, which is expected since the high DOS both increases the strength of screening and the weighting in the gap equation. Indeed, the dependence of $\lambda$ along the $\mu$-axis echoes the salient features of the DOS (Fig.~\ref{fig:DOS}). The solution is predominantly the 2D irreducible representation corresponding to a $p$-wave gap function (Fig.~\ref{fig:RPA_main}e), though a non-degenerate extended $s$-wave solution --- where the order parameter changes sign between the inner and outer parts of the Fermi surface ---  is competitive in a narrow sliver of doping at the vHS, especially for larger $\epsilon_r$ \cite{Supplement}.

Moving to finite fields, we find that the $T_c$ peak in Fig.~\ref{fig:RPA_main}a splits off into two branches which follow the vHS of the majority and minority spins. The spin projection involved in pairing remains at the van Hove filling, while the detuning of the opposite `spectator' spin leads to a change in screening properties and hence $\lambda_p$. Na\"ively, shifting the spectator spin away from the vHS would sharply reduce the DOS and suppress KL superconductivity. However, owing to the narrow dispersion, a small Zeeman shift significantly changes the Fermi surface, and hence the polarization function. Along the minority branch, the spectator Fermi surface expands and fills in the voids at the Dirac momenta, leading to a slight enhancement of $\Pi(q_{P})$ (right panel of Fig.~\ref{fig:RPA_main}d). On the other hand for the majority branch, the Fermi surface shrinks into small pockets such that screening at $q_{P}$, which is deleterious to the superconductivity, is less effective (middle panel of Fig.~\ref{fig:RPA_main}d). [This saturates when the field fully polarizes the spins, which occurs at $B\gtrsim10\,\text{T}$ for our parameters.] This therefore leads to a strongly asymmetric contribution from the first term in Eq.~\ref{eqn:lambda_p} and hence stronger pairing in the `majority branch' (doping towards CN).

In the case of a physical magnetic field, even though $\bm{B}$ is actually applied parallel to the graphene sheets, the orbital coupling (enabled by the finite interlayer distance) may be non-negligible owing to the small energy scales involved. For $B\sim 1\,\text{T}$, the typical  depairing energy $\epsilon_K(\bm{k})-\epsilon_{K'}(-\bm{k})$ is of order $0.01\,$meV, which is comparable to $T_c$. Indeed, upon incorporating the orbital effects of the magnetic field in the gap equation \cite{Supplement}, we find a substantial suppression in $\lambda$ which may lead to a peak in $T_c$ at a finite $B$ (Fig.~\ref{fig:RPA_main}b). This effect is absent if the flavors are imbalanced instead by Ising spin-orbit coupling.

\begin{figure*}
    \centering
    \includegraphics[width=\textwidth]{main_panel.pdf}
    \caption{\textbf{Superconductivity in a functional renormalization group calculation}. (a) Patching scheme of the majority spin Fermi surface with $N_p=60$ patches for a chemical potential $0.05$meV above the vHS. (b,c) BCS vertex $\Gamma(\mathbf{k},\mathbf{k}')$ at $T=10^4$K, $12$K respectively. Note that the value of $\Gamma(\mathbf{k},\mathbf{k}')$ depends on the size of the patches at $\mathbf{k}$ and $\mathbf{k'}$. The numbers refer to the patch labels defined in (a). We use the Coulomb interaction with $d_\textrm{sc}=25$nm and $\epsilon_r=17.5$. (d) The divergence of the most negative eigenvalue of the superconducting susceptibility signifies the superconducting $T_c$. (e) The two-fold degenerate leading eigenvector of the gap equation is a $p$-wave order parameter. We show the $p_x$ solution. (f) The flow of two components of the vertex representative of intra-pocket ($U=\Gamma(20,20)$) and inter-pocket ($V=\Gamma(19,20)$) scattering. The polarization bubbles are evaluated with $N_\textrm{ref}=101$.}
    \label{fig:dep}
\end{figure*}

\textit{FRG.---} We perform an FRG calculation in order to confirm that the superconductivity persists when fluctuations beyond the RPA are taken into account. 
FRG is a intermediate-coupling approach that involves integrating out high-energy degrees of freedom, in order to obtain a renormalized interaction valid close to the Fermi surface \cite{POLCHINSKI1984269,Shankar,metzner2012RMP,Platt2013review,Salmhofer2019review,Dupuis2021review,Kopietz2010book}. FRG has been used to study  SC in both TBG \cite{kennes2018did,klebl2020frg,tang2019fwave,classen2019triangular}
and RTG \cite{qin2022FRG}, yet has not to date been applied to biased BBG. The central object of the FRG calculation is the temperature-dependent 4-point vertex $\gamma_{ab;cd}(\bp_1,\bp_2,\bp_3)$, where the composite subscripts $a,b,\ldots$ label both spin and valley. The FRG equations are \cite{Platt2013review}
\begin{align}
\label{eq:FRG_flow}
    &\dot\gamma_{abcd}(\bp_1,\bp_2,\bp_3)=\int_{\bk}\sum_{xy}\frac{1}{2}\times \\ &\bigg[-\dot\pi^{pp}_{xy}(\bk,\bp_1+\bp_2)
    \gamma_{abxy}(\bp_1,\bp_2,\bk)\gamma^{ *}_{cdxy}(\bp_3,\bp_4,\bk)\nonumber
   \\&+2\dot\pi^{ph}_{xy}(\bk,\bp_1-\bp_3)\gamma^{*}_{cxay}(\bp_3,\bk,\bp_1)\gamma_{bxdy}(\bp_2,\bk,\bp_4)\nonumber
   \\&-2\dot\pi^{ph}_{xy}(\bk,\bp_2-\bp_3)\gamma^{*}_{cxby}(\bp_3,\bk,\bp_2)\gamma_{axdy}(\bp_1,\bk,\bp_4)\bigg]\nonumber,
\end{align}
where $\dot\ \equiv\partial_T$ (we employ the temperature-flow FRG scheme \cite{Platt2013review}, where the RG scale is set by the temperature). The momentum arguments $\bp_i$ of the 4-point vertex are chosen to be $N_p$ equally spaced patch momenta on the Fermi surface (Fig.~\ref{fig:dep}a). To capture the details of the bandstructure, we evaluate the polarization bubbles $\pi^{pp}$ and $\pi^{ph}$ on a fine $N_\textrm{ref}\times N_\textrm{ref}$ mesh~\cite{Supplement}. We start the FRG flow at a temperature of $10^4$K and flow down to $10^{-3}$K in logarithmic steps. The temperature at which the superconducting susceptibility diverges defines the superconducting critical temperature $T_c$.

The Coulomb interaction is a monotonically decreasing function of momentum transfer and this sets the structure of the initial vertex: The maximum values of the vertex are attained for small intra-pocket momentum transfers or for small momentum transfers between points in different pockets close to the $K$-point (Fig.~\ref{fig:dep}b). The gap matrix in Fig.~\ref{fig:dep}b echoes the block structure of Eq.~\ref{eq:M}. We have $U>V$ and no superconducting instability. At lower temperatures, the bare interaction has been screened such that the components of the vertex with larger momentum transfers are larger than those with small momentum transfer (Fig.~\ref{fig:dep}c), i.e.~$V>U$ in Eq.~\eqref{eq:M}. $U$ is screened more heavily than $V$ since $\Pi(0)>\Pi(q_{P})$.  In Fig.~\ref{fig:dep}f we show the FRG flow of two representative components of the vertex function that show this screening behaviour as in Eq.~\eqref{eqn:lambda_p}.  At the end of the FRG flow we have $V>U$ which leads to a divergence in the most negative eigenvalue of the superconducting susceptibility (Fig.~\ref{fig:dep}d) in the $p$-wave channel (Fig.~\ref{fig:dep}e) in agreement with both the pocket model and the RPA. However, in contrast to the RPA calculation, screening in the FRG can lead to $U<0$, further enhancing the superconductivity. For the parameters $d_\textrm{sc}=25$nm and $\epsilon_r=17.5$ we find $T_c\approx0.5$K, though since this is a nodal order parameter, disorder would reduce this scale.

\textit{Conclusions}.---We have shown that a Kohn-Luttinger mechanism based on overscreening of an initially purely repulsive interaction provides a unified explanation for superconductivity in BBG either in the presence of a parallel magnetic field or SOC induced by proximity to WSe$_2$. The Kohn-Luttinger mechanism for parabolic bands in two dimensions is weak~\cite{Ghazaryan2021,RaghuKivelson,OU3}, however the deviations from parabolicity (`trigonal warping') in BBG as well as the imbalancing of the occupation numbers of the different isospin flavours due to an effective Zeeman field enhance the effect. Furthermore, the flat bands of BBG induced by the applied displacement field as well as the proximity to a van Hove singularity provide a high density of states, leading to a $T_c$ in a realistic range.  We find robust $p$-wave superconductivity in both  RPA and  FRG calculations, which lends support to a simplified three-pocket model. Intra-pocket interactions are more heavily screened than inter-pocket interactions, leading to an overall attraction in the $p$-wave channel. The details of the screening lead to enhanced superconductivity when doping towards charge neutrality as opposed to away from charge neutrality which is consistent with experimental observations. Extending the existing STM studies on BBG to measure Andreev reflection~\cite{Andreev} or performing quasi particle interference experiments \cite{Pangburn} could provide an experimental test to confirm the $p$-wave nature of the superconducting order parameter. 

The Fermi surface of RTG with trigonal warping and an applied displacement field also consists of three separate pockets for a range of doping close to the van Hove singularity and therefore the Kohn-Luttinger mechanism described by our three-pocket model would likely result in superconductivity in that material too, as has been observed in experiments \cite{RTG_SC}. Indeed Ref.~\cite{BBG_KL} showed that a Kohn-Luttinger mechanism can explain superconductivity in both BBG and RTG. The three-pocket model thus provides a unifying explanation for superconductivity in graphene multilayers, unlike theories of superconductivity in RTG based on the annular Fermi surface \cite{Ghazaryan2021,qin2022FRG} (although Ref.~\cite{Ghazaryan2021} also looked at the three-pocket regime).  On the other hand, for twisted bilayer graphene, experiments \cite{Sharpe_2019,Serlin900,Lu2019,Cao2018,Cao2018b,Yankowitz_2019,Park_2021,Stepanov_2020,Wu_2021,Zondiner_2020,uri2020mapping,saito2020independent} as well as numerics \cite{IKS_PRL,kang2021cascades} show a single simply-connected Fermi surface per flavour, such that a different mechanism must be responsible for superconductivity, underscoring the different physics at play in moir\'eless vs.~moir\'e graphene multilayers \cite{Patri2022}.

\begin{acknowledgements}
\textit{Acknowledgements}.--- We thank the authors of \cite{BBG_KL} for valuable comments on an earlier version of this manuscript. 
We acknowledge funding from the European Research Council (ERC) under the European Union’s Horizon 2020 research and innovation program via ERC-StG-Neupert-757867-PARATOP (GW) and ERC-StG-Parameswaran-804213-TMCS (YHK, SAP), the Royal Society via a University Research Fellowship (NB), and EPSRC Grant EP/S020527/1
 (SHS). Statement of compliance with EPSRC policy framework
on research data: This publication is theoretical work
that does not require supporting research data.
\end{acknowledgements}

\bibliography{bib}

\newpage
\clearpage

\begin{appendix}
\onecolumngrid
	\begin{center}
		\textbf{\large --- Supplementary Material ---\\ Superconductivity from repulsive interactions in Bernal-stacked bilayer graphene}\\
		\medskip
		\text{Glenn Wagner, Yves H. Kwan, Nick Bultinck, Steven H. Simon and S.A.~Parameswaran}
	\end{center}
	
		\setcounter{equation}{0}
	\setcounter{figure}{0}
	\setcounter{table}{0}
	\setcounter{page}{1}
	\makeatletter
	\renewcommand{\theequation}{S\arabic{equation}}
	\renewcommand{\thefigure}{S\arabic{figure}}
	\renewcommand{\bibnumfmt}[1]{[S#1]}

\section{Single-particle Hamiltonian}
We use the single-particle Hamiltonian from Ref.~\cite{Jung2014}. We include both the Zeeman and the orbital effects of an in-plane magnetic field $\vec B=B(\cos\theta,\sin\theta,0)$ with corresponding gauge field $\vec A=z\vec B\times \vec{\hat z}$. In the basis $\{1A,1B,2A,2B\}$ the Hamiltonian in valley $\tau=\pm1$ is
\begin{equation}
\label{eq:Ham_app}
H_\tau(\mathbf{k})=\left(\begin{array}{cccc}
D/2+\lambda_I\tau s_z/2-\mu & v \pi^{\dagger}_1 & -v_{4} \pi^{\dagger} & -v_{3} \pi \\
v \pi_1 & \Delta^{\prime}+D/2+\lambda_I\tau s_z/2-\mu & t_{1} & -v_{4} \pi^{\dagger} \\
-v_{4} \pi & t_{1} & \Delta^{\prime}-D/2-\mu & v \pi^{\dagger}_2 \\
-v_{3} \pi^{\dagger} & -v_{4} \pi & v \pi_2 & -D/2-\mu
\end{array}\right)+\mu_B \vec B\cdot\vec s,
\end{equation}
where $D$ is the displacement field ($|D|=50$meV in Ref.~\cite{SC_BLG}) and $\pi=\hbar(\tau k_x+ik_y)$. The minimally coupled momenta $\pi_{1,2}$ in the two layers are obtained via $\vec k\to \vec k\pm \frac{eBd}{2\hbar}(\sin\theta,-\cos\theta)$, where the upper (lower) sign refers to layer 1 (2). $d=0.35$nm is the distance between the graphene layers.
The velocities are defined via $v=t_{0} \sqrt{3} a / 2 \hbar, v_{3}=t_{3} \sqrt{3} a / 2 \hbar$ and $v_{4}=t_{4} \sqrt{3} a / 2 \hbar$ where $a$ is the lattice constant of graphene. $s_i$ are the Pauli matrices associated with spin. We note that the Ising SOC only couples to layer 1, which is assumed to be the layer in contact with the SOC-inducing WSe$_2$. It is clear from the experimental data that the Ising SOC only affects the layer in contact with the WSe$_2$ since the SC is only seen when the sign of the displacement field is such that the electrons are polarized towards the WSe$_2$. For the opposite sign of the displacement field, the electrons are polarized in the opposite layer and hence do not feel the effect of the WSe$_2$ and SC is not seen. The bands close to the Fermi surface are already strongly layer-polarized for moderate displacement fields of $D>20$meV. Since we consider hole-doped BBG (as studied in the experiments), we focus on the valence band of \eqref{eq:Ham_app} and neglect the conduction band which is separated by a gap.
\begin{table}[h]
\centering
\caption{Tight-binding parameters from Ref.~\cite{Jung2014}}
\begin{tabular}{p{2.5cm} p{1.3cm} p{1.3cm} p{1.3cm} p{1.3cm} p{1.3cm}} 
\toprule
Parameter & $t_0$ & $t_1$ & $t_3$ & $t_4$ & $\Delta'$ \\ \hline
Value [eV] & $2.61$ & $0.361$ & $0.283$ & $0.138$ & $0.015$ \\
\label{tab:HoppingParameters}
\end{tabular}
\end{table}

\section{Additional details of RPA calculation}
To gain insight into how a Kohn-Luttinger-like mechanism can lead to superconductivity in BBG, we perform a numerical analysis in the random phase approximation (RPA). The bare gate-screened interaction $V^0(\bm{q})=\frac{e^2}{2\epsilon_0\epsilon q}\tanh qd_\text{sc}$, with relative permittivity $\epsilon$ and gate distance $d_\text{sc}$, is purely repulsive. Accounting for electronic screening in the RPA, this leads to the renormalized interaction
\begin{equation}
    V^\text{RPA}(\bm{q})=\frac{V^0(\bm{q})}{1+\Pi(\bm{q})V^0(\bm{q})},
\end{equation}
where we have defined the static polarization function $\Pi(\bm{q})=\sum_f \Pi_f(\bm{q})$, summed over flavors $f$
\begin{equation}\label{eqn:polarization}
    \Pi_f(\bm{q})=\frac{1}{A}\sum_{\bm{k}}\frac{n_F(\epsilon_f(\bm{k}+\bm{q}))-n_F(\epsilon_f(\bm{k}))}{\epsilon_f(\bm{k})-\epsilon_f(\bm{k}+\bm{q})}.
\end{equation}
Note that while each flavor experiences the same screened interaction, the polarization above can differ depending on external fields and perturbations. We have neglected the Bloch form factors for simplicity---we anticipate that these only affect the physics quantitatively for the following reasons: First, at low doping only small momentum scattering is important, and the magnitude of the form factors $\langle u_{\vec{k}-\vec{q}}|u_{\vec{k}}\rangle$ does not vary significantly for such small $\vec{q}$. Secondly, the phases of the form factors do not appear in the polarization $\Pi(\vec{q})$, and in a suitable gauge they also cancel in the part of the interaction which describes scattering of zero-momentum electron pairs due to time-reversal symmetry (this is the only part of the interaction which enters the gap equation). To compute the screened interaction, we compute the non-interacting dispersion $\epsilon_f(\bm{k})$ on a uniform triangular grid of approximately $1000\times1000$ points. The momentum extent of the grid is $>3.5\%$ of the reciprocal lattice vector---larger than two times the maximum separation of points on the Fermi surface. The polarization function (Eq.~\ref{eqn:polarization}) is computed on a coarser grid of $>400\times400$ points. The Fermi functions are evaluated at $T=0$ with a small infrared cutoff in the denominator of $\Pi_f(\bm{q})$ to avoid divergences. The results are robust to including a small $T$ in the occupation factors. 

\begin{figure}
    \centering
    \includegraphics[width=0.5\columnwidth]{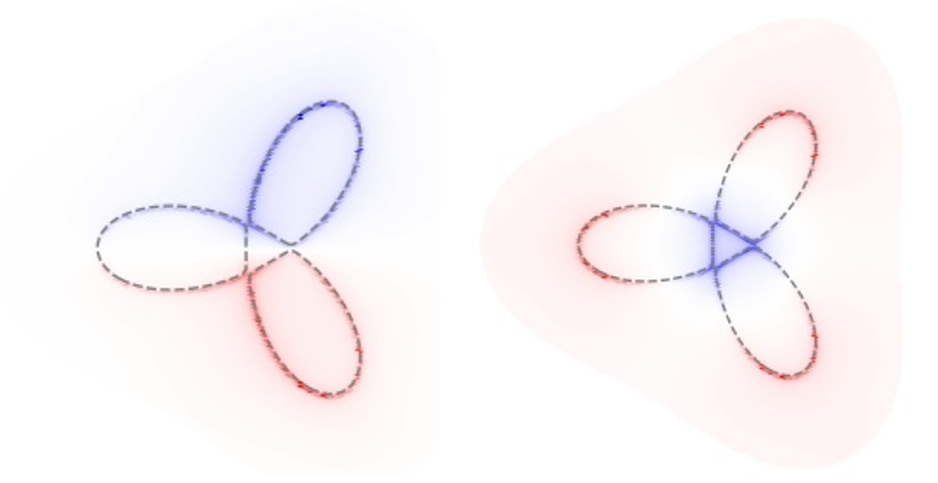}
    \caption{\textbf{Example RPA gap functions in the grid scheme.}}
    \label{fig:grid_s_vs_p}
\end{figure}

It is possible to proceed by solving the gap function on the same grid. Assuming intervalley pairing between up-spins with vanishing pair momentum, the gap matrix is
\begin{equation}
    M(\bm{p},\bm{p}')=-\frac{1}{A}\sqrt{\pi^\text{pp}(\bm{p})}V^\text{RPA}(\bm{p},\bm{p}')\sqrt{\pi^\text{pp}(\bm{p}')}
\end{equation}
where we have defined the particle-particle susceptibility
\begin{equation}\label{eqn:pipp}
    \pi^\text{pp}(\bm{p})=\frac{1-n_F(\epsilon_{+\uparrow}(\bm{p}))-n_F(\epsilon_{-\uparrow}(-\bm{p}))}{\epsilon_{+\uparrow}(\bm{p})+\epsilon_{-\uparrow}(-\bm{p})}.
\end{equation}
[Recall that $\uparrow,\downarrow$ should be understood as accounting for spin-valley locking in the case of Ising SOC]. To reduce the size of $M$, only grid points whose energy is within a UV cutoff of the Fermi energy are kept. $M$ is diagonalized at different temperatures until an eigenvalue $1$ is reached, signalling $T_c$. Representative gap functions are shown in Fig.~\ref{fig:grid_s_vs_p}. While this procedure is straightforward at higher temperatures, the calculation becomes increasingly difficult at lower temperatures. The reason is that the relevant regions of momentum space involved in pairing form an increasingly narrow sleeve around the Fermi surface. A dense grid is required to capture the Fermi surface properly and prevent discretization artifacts.

Therefore, we employ a patch scheme that sidesteps this issue directly. Each Fermi contour is divided into segments of equal length $\delta_{\bm{p}}<10^{-4}$ of the reciprocal lattice vector indexed by momentum $\bm{p}$. No fewer than $25$ patches are included on each Fermi contour. $V^\text{RPA}$, which is still defined on the grid, is interpolated to determine the interaction between different patches. The dispersion in the direction perpendicular to the Fermi surface is treated in the linear approximation $\epsilon_f(\bm{p},k)=\hbar v_{\bm{p}} k$ and integrated (with UV cutoff $E_0$) to obtain the logarthmic Cooper divergence. The resulting gap equation is characterized by the symmetrized gap matrix
\begin{equation}\label{eqn:gap_matrix}
    M_{\bm{p},\bm{p}'}=-\frac{1}{(2\pi)^2\hbar}\sqrt{\frac{\delta_{\bm{p}}\delta_{\bm{p}'}}{v_{\bm{p}}v_{\bm{p}'}}}V^{\text{RPA}}(\bm{p}-\bm{p}'),
\end{equation}
whose largest positive eigenvalue $\lambda$ corresponds to a superconducting solution with $T_c\sim E_0e^{-\frac{1}{\lambda}}$.
Eq.~\ref{eqn:gap_matrix} has to be diagonalized just once since we are implicitly working in the logarithmic temperature scaling regime. We note that the grid and patch scheme yield qualitatively similar results in e.g.~the competition between $p$-wave and extended $s$-wave solutions, and the asymmetry between the minority and majority branches.

\begin{figure*}
    \centering
    \includegraphics[width=\textwidth]{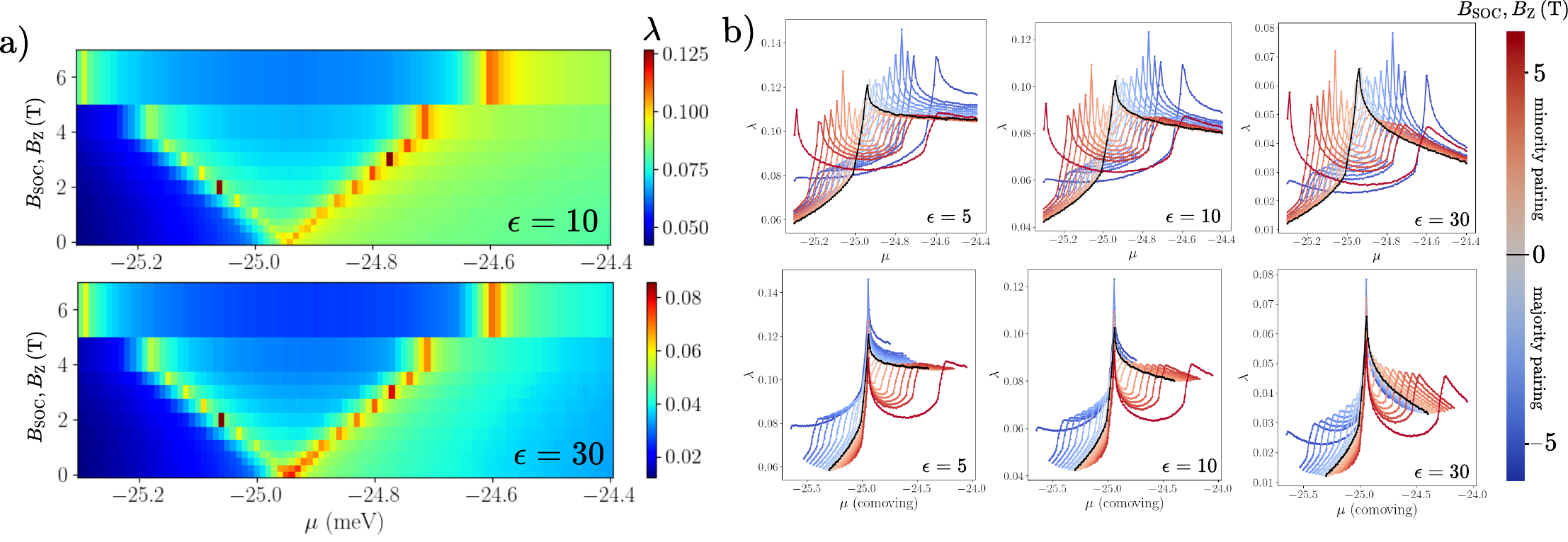}
    \caption{\textbf{Superconductivity in RPA at different interaction strengths.} \textbf{a)} Maximum gap matrix eigenvalue $\lambda$ as a function of applied Zeeman field $B$ and chemical potential $\mu$. Same as Fig.~\ref{fig:RPA_main} in main text but with different relative permittivity $\epsilon$. \textbf{b)} Results for different Zeeman field strengths unfolded as individual line plots and color coded. $B=0$ is plotted in black. For positive (negative) $B$, only pairing in the minority (majority spin species is shown. The second row shifts the plots horizontally to the comoving frame, such that the vHS's of the spin species that is allowed to pair all line up.}
    \label{fig:extra_RPA}
\end{figure*}

In Fig.~\ref{fig:extra_RPA}, we show additional results of the RPA calculation of superconductivity. In Fig.~\ref{fig:extra_RPA}a, we provide color plots of $\lambda$ in the $B-\mu$ plane for weaker interaction strengths. Note that the superconductor at $B=0$ becomes comparatively stronger for larger $\epsilon$. In Fig.~\ref{fig:extra_RPA}b, we unfold the plots for different $B$ to allow for easier comparison. Furthermore, we distinguish between pairing between majority spins (negative $B$) and between minority spins (positive $B$). Therefore, the majority branch referred to in the main text corresponds to negative fields here. In the second row, we further shift the lines horizontally by the Zeeman energy so the van Hove singularity points are all coincident. We point out several features. For strong interactions and majority pairing, the tail away from the vHS towards neutrality still has an appreciable $\lambda$. For both spin species, there is a small `shadow' peak in $\lambda$ whenever the spectator spin is at its vHS. This effect is stronger for weaker interactions, consistent with the stronger role played by $q=0$ screening as $\epsilon$ is increased. There appear to be some modulations in the peak height as $B$ is varied---this is partially caused by sampling effects from the discrete grid of $\mu$ points that the calculations were performed on.

\begin{figure}
    \centering
    \includegraphics[width=0.3\columnwidth]{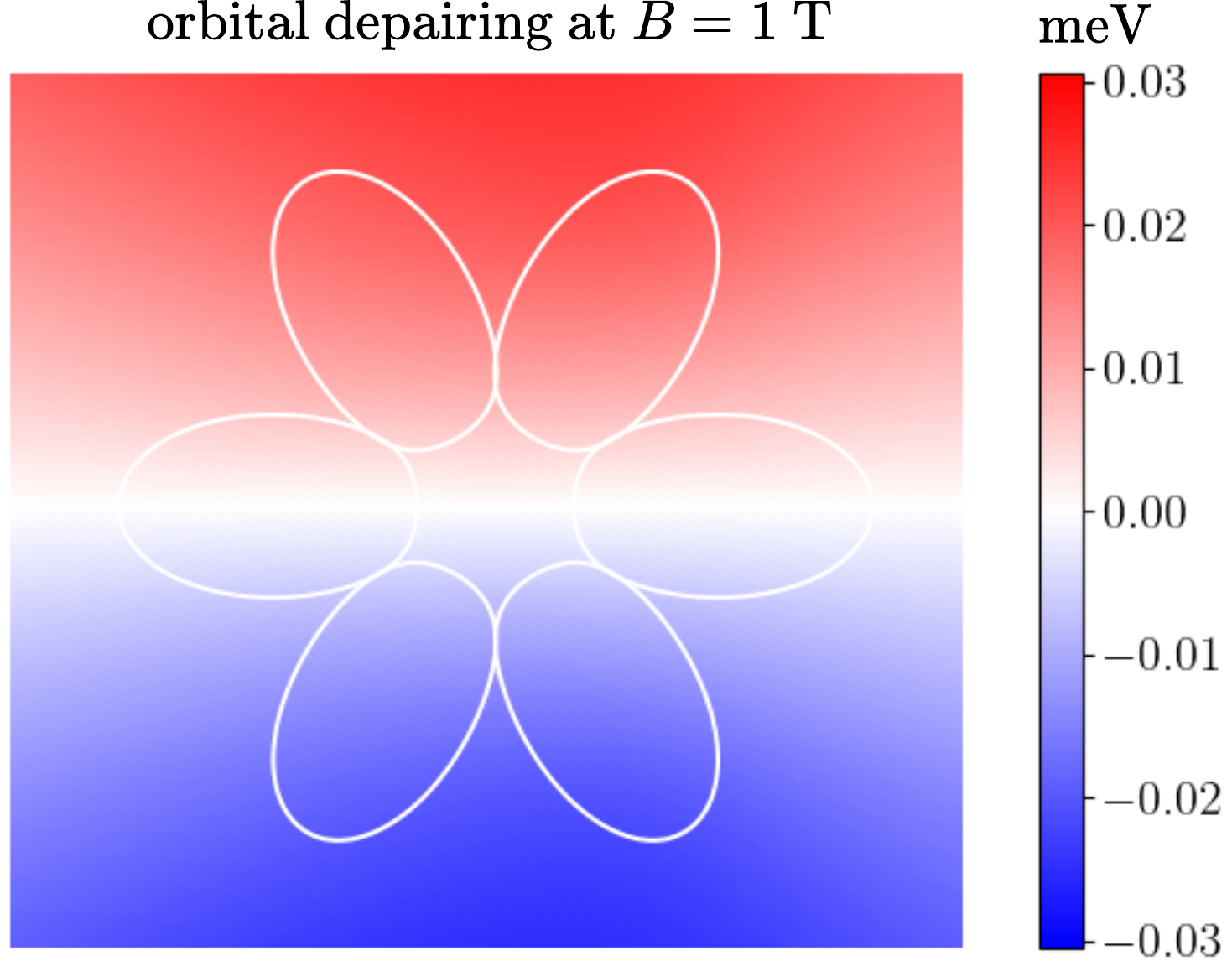}
    \caption{\textbf{Orbital depairing due to in-plane magnetic field.} The Fermi contours for both valleys at $\mu=-24.7\,$meV are shown for reference. Magnetic field of strength $B=1$\,T, oriented along the $\hat{x}$-axis.}
    \label{fig:orbital_depairing}
\end{figure}

When studying the orbital effects of an in-plane magnetic field, we assume that the dominant effect is the energy depairing of the Cooper pairs. Therefore we neglect the changes to $V^\text{RPA}(\bm{q})$. Consider two time-reversal related momenta formerly on the Fermi surface (after Zeeman shifts have been included). After accounting for orbital coupling, their energies can be decomposed into symmetric and antisymmetric contributions
\begin{equation}
    \epsilon_\pm(\pm\bm{p})=\pm a(\bm{p})+b(\bm{p}).
\end{equation}
For BBG, the symmetric shift $b(\bm{p})$ is more than an order of magnitude smaller than the the antisymmetric part $a(\bm{p})$. Hence we ignore $b(\bm{p})$, and retain the same patching as in the $B=0$ case. This neglects the possibility pairing along a slightly different momentum contour, as well as non-zero momentum pairing, so our results should be viewed as an overestimate of the deleterious effects of orbital coupling. Fig.~\ref{fig:orbital_depairing} shows the typical scale of depairing.

To account for the antisymmetric shift of energies in the gap equation, we recast particle-particle susceptibility $\pi^\text{pp}(\epsilon,a)$ [Eq.~\ref{eqn:pipp}] into its zero-field part (which still contains the logarithmic divergence) and a correction
\begin{equation}
    \pi^\text{pp}(\epsilon,a)=\frac{1}{2\epsilon}\tanh\frac{\beta\epsilon}{2}+\frac{1}{2\epsilon}[2n_F(\epsilon)-n_F(\epsilon-a)-n_F(\epsilon+a)]
\end{equation}
where $\epsilon=\hbar v k$. Integrating over the radial momentum $k$ leads to
\begin{equation}
    \frac{1}{\hbar v}\left[\ln\frac{E_0}{k_B T}+\int_{-\infty}^\infty \frac{dx}{2x}\,\left(\frac{2}{e^x+1}-\frac{1}{e^{x-\beta a}+1}-\frac{1}{e^{x+\beta a}+1}\right)\right].
\end{equation}
The integral above is convergent and can be evaluated numerically for each patch. Since we are no longer in the scaling regime, the gap matrix has to be solved for different $T$ until an eigenvalue of $1$ is reached.

\section{Extended $s$-wave hotspot solution}

\begin{figure}
    \centering
    \includegraphics[width=0.25\columnwidth]{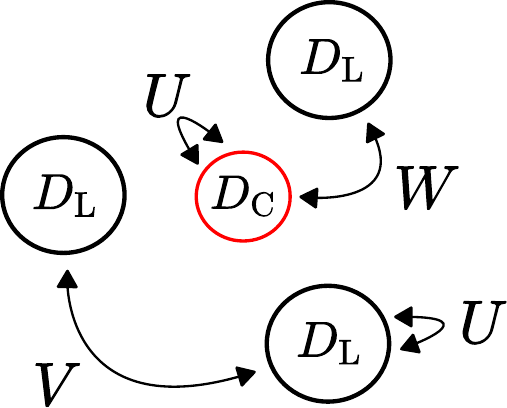}
    \caption{\textbf{Pocket model with four patches, including a hotspot at the Dirac momentum.}}
    \label{fig:fourhotspot}
\end{figure}

In this section, in order to understand the competing extended $s$-wave solution (Fig.~\ref{fig:grid_s_vs_p}, right), we discuss an extended version of the pocket model (four patch model) which, in addition to the three lobes, now includes a central hotspot around the Dirac momentum (Fig.~\ref{fig:fourhotspot}). This requires a new coupling $W$ between the central pocket and the lobes. We also need to account for different weighting factors $D_{\text{C}},D_{\text{L}}\sim\sqrt{\frac{\delta_{\bm{p}}}{v_{\bm{p}}}}$ in the gap equation (Eq.~\ref{eqn:gap_matrix}), which depend on the local DOS and the pocket geometry.

The gap matrix in this four patch model reads
\begin{equation}
    M=-\begin{pmatrix}
    D^2& DW&DW&DW\\
    DW&1&V&V\\
    DW&V&1&V\\
    DW&V&V&1
    \end{pmatrix},
\end{equation}
where we have defined $D=D_\text{C}$, and measured quantities in units of $D_\text{L}$ and $U$ for simplicity. The $p$-wave solution is 
\begin{equation}
    \Delta_p\sim\begin{pmatrix}
    0\\1\\ e^{\pm\frac{2\pi i}{3}}\\
     e^{\pm\frac{4\pi i}{3}}
    \end{pmatrix},\quad \lambda_p=V-1.
\end{equation}
To construct the extended $s$-wave solution, we consider the basis $\phi_\alpha=[1,0,0,0]^T,\phi_\beta=\frac{1}{\sqrt{3}}[0,1,1,1]^T$, leading to the effective gap matrix
\begin{equation}
    M_{\tilde{s}}=-\begin{pmatrix}
    D^2 & \sqrt{3}DW\\
    \sqrt{3}DW&1+2V
    \end{pmatrix}.
\end{equation}
Since we are interested in the most positive eigenvalue, the eigenvector has coefficients of opposite sign for the two basis elements. The eigenvalue is
\begin{equation}
    \lambda_{\tilde{s}}=-\left(\frac{D^2+1+2V}{2}\right)+\sqrt{\left(\frac{D^2-1-2V}{2}\right)^2+3D^2W^2}.
\end{equation}
We are interested in determining the phase diagram as a function of $V,D,W$. The $p$-wave solution is physical as long as $V>1$. Hence, it remains to check the condition $\lambda_{\tilde{s}}>\max{(0,\lambda_p)}$. 

\begin{figure}
    \centering
    \includegraphics[width=0.6\columnwidth]{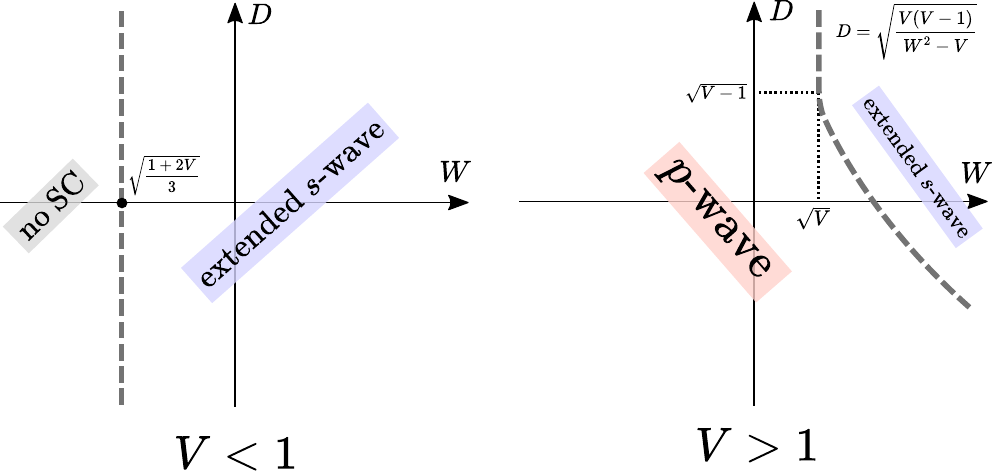}
    \caption{\textbf{Phase diagram of four patch model.} Axes are on a logarithmic scale.}
    \label{fig:s_vs_p}
\end{figure}

Consider first $V<1$, where the relevant bound is $\lambda_{\tilde{s}}>0$. This leads to the $D$-independent condition 
\begin{equation}
W>\sqrt{\frac{1+2V}{3}},\quad (V<1).
\end{equation}
For $V>1$, we need to consider when $\lambda_{\tilde{s}}>\lambda_p$, leading to
\begin{equation}
    W>\sqrt{V+\frac{V(V-1)}{D^2}},\quad (V>1),
\end{equation}
which asymptotes to $W>\sqrt{V}$ as $D\rightarrow\infty$. Hence, extended $s$-wave pairing is not guaranteed even in this limit. The phase diagram is summarized in Fig.~\ref{fig:s_vs_p}.

For reasonable interaction strengths in BBG, we expect $V>1$ after screening and therefore competition between the two superconducting orders. It is plausible that the condition $W>\sqrt V$ is met, but given the strong screening at $q\simeq 0$ (so that $V\gg 1$), the requirement on $D$ can be quite severe, explaining the restriction of extended $s$-wave pairing to a narrow density sliver at the van Hove filling. 

The competition as a function of $\epsilon$ at fixed $\mu$ can also be rationalized in this simple model. We assume that the interactions are sufficiently strong that $U$ is $\epsilon$-independent. The threshhold ratio for extended $s$-wave pairing is then $\frac{W}{\sqrt{V}}\sim 1$. In obvious notation, this is given by
\begin{equation}
    \frac{W}{\sqrt{V}}=\frac{f(W)}{f(V)}\frac{\sqrt{\epsilon+\Pi(V)f(V)}}{\epsilon+\Pi(W)f(W)}
\end{equation}
where $f(q)$ and $\Pi(q)$ are the bare interaction (with the relative permittivity removed) and static polarization function defined in the main text. For large $\epsilon$, this behaves as $\sim\sqrt{\epsilon}$. For small $\epsilon$, the leading dependence is $\sim\epsilon[\Pi(W)f(W)-2\Pi(V)f(V)]$. Given that $f(q)$ and $\Pi(q)$ are typically monotonic decreasing positive functions, $\Pi(q)$ decreases dramatically from $q=0$, and $\frac{q_W}{q_V}\lesssim 0.5$, the quantity in square brackets is likely positive. Therefore, $\frac{W}{\sqrt{V}}$ is expected to be a decreasing function of interaction strength, meaning that $p$-wave pairing is increasingly favored. Finally, we note that the competition between the $s$-wave and the $p$-wave solution may also be tuned by the tight-binding parameters of the Hamiltonian \cite{PrivComm}.

\section{Initial FRG vertex}
For the initial vertex which is fed into the FRG equation we use the Coulomb interaction projected into the valence band subspace of \eqref{eq:Ham}. The dual gate screened interaction is 
\begin{equation}
    V(q)=\frac{e^2}{2\epsilon_0\epsilon_r q}\tanh{qd}.
\end{equation}
We define the intra- and intervalley Coulomb interaction via 
\begin{equation}
    v_{\tau\tau'}(q)=
\begin{cases}
\frac{1}{A}V(q) \textrm{ for } \tau=\tau',\\
\frac{1}{A}V(Q) \textrm{ for } \tau\neq\tau'
\end{cases}
\end{equation}
where $Q=4\pi/(3a)$ is the intervalley momentum ($a$ is the lattice constant of graphene) and $A$ is the total area of the system. The matrix elements of the Coulomb interaction are 
\begin{align}
    V_{\tau_1\sigma_1\tau_2\sigma_2\tau_3\sigma_3\tau_4\sigma_4}(\bp_1,\bp_2,\bp_3)&=v_{\tau_1\tau_3}(\bp_3-\bp_1)\phi_{\alpha\tau_1\sigma_1}^*(\bp_1)\phi_{\beta\tau_2\sigma_2}^*(\bp_2)\phi_{\alpha\tau_3\sigma_3}(\bp_3)\phi_{\beta\tau_4\sigma_4}(\bp_1+\bp_2-\bp_3),
\end{align}
where $\phi_{\alpha\tau\sigma}(\bp)$ is the eigenvector of \eqref{eq:Ham} corresponding to the valence band and $\alpha, \beta$ are sublattice and layer indices which are summed over. We assume the Coulomb interaction is layer-independent, which is a good approximation since for a large displacement field the valence band is highly layer-polarized close to the Fermi energy. The 4-point vertex is then obtained by anti-symmetrization
\begin{equation}
    \gamma^{\uparrow\uparrow\uparrow\uparrow}_{\tau_1\tau_2\tau_3\tau_4}(\bp_1,\bp_2,\bp_3)=V_{\tau_1\uparrow\tau_2\uparrow\tau_3\uparrow\tau_4\uparrow}(\bp_1,\bp_2,\bp_3)-V_{\tau_2\uparrow\tau_1\uparrow\tau_3\uparrow\tau_4\uparrow}(\bp_2,\bp_1,\bp_3),
\end{equation}
\begin{equation}
    \gamma^{\downarrow\downarrow\downarrow\downarrow}_{\tau_1\tau_2\tau_3\tau_4}(\bp_1,\bp_2,\bp_3)=V_{\tau_1\downarrow\tau_2\downarrow\tau_3\downarrow\tau_4\downarrow}(\bp_1,\bp_2,\bp_3)-V_{\tau_2\downarrow\tau_1\downarrow\tau_3\downarrow\tau_4\downarrow}(\bp_2,\bp_1,\bp_3),
\end{equation}
\begin{equation}
    \gamma^{\uparrow\downarrow\uparrow\downarrow}_{\tau_1\tau_2\tau_3\tau_4}(\bp_1,\bp_2,\bp_3)=V_{\tau_1\uparrow\tau_2\downarrow\tau_3\uparrow\tau_4\downarrow}(\bp_1,\bp_2,\bp_3).
\end{equation}

\section{FRG equations}
The central object of the FRG calculation is the scale-dependent 4-point vertex $\gamma_{ab;cd}(\bp_1,\bp_2,\bp_3)$, where in our case the RG scale parameter is the temperature $T$. The subscripts $a,b,\ldots$ label spin and valley. The FRG equations without assuming spin-SU(2) invariance are \cite{Platt2013review}
\begin{align}
\label{eq:FRG_flow_app}
    \dot\gamma_{ab;cd}(\bp_1,\bp_2,\bp_3)  &=\frac{1}{2}\int_{\bk,xy}\bigg[-\dot\pi^{pp}_{xy}(\bk,\bp_1+\bp_2)
    \gamma_{ab;xy}(\bp_1,\bp_2,\bk)\gamma^{ *}_{cd;xy}(\bp_3,\bp_4,\bk)\nonumber
   \\&\qquad\qquad+2\dot\pi^{ph}_{xy}(\bk,-\bp_3+\bp_1)\gamma^{*}_{cx;ay}(\bp_3,\bk,\bp_1)\gamma_{bx;dy}(\bp_2,\bk,\bp_4)
   \\&\qquad\qquad-2\dot\pi^{ph}_{xy}(\bk,-\bp_3+\bp_2)\gamma^{*}_{cx;by}(\bp_3,\bk,\bp_2)\gamma_{ax;dy}(\bp_1,\bk,\bp_4)\bigg]\nonumber,
\end{align}
where $\dot\ \equiv\partial_T$. The polarization bubbles are given by
\begin{gather}
    \pi^{ph}_{xy}(\bk,\bq)=\frac{n_F\left(\epsilon_x(\bk)\right)-n_F\left(\epsilon_y(\bk-\bq)\right)}{\epsilon_x(\bk)-\epsilon_y(\bk-\bq)},\\
    \pi^{pp}_{xy}(\bk,\bq)=\frac{1-n_F\left(\epsilon_x(\bk)\right)-n_F\left(\epsilon_y(\bq-\bk)\right)}{\epsilon_x(\bk)+\epsilon_y(\bq-\bk)},
\end{gather}
where $\epsilon_x(\bk)$ are the single-particle energies of the Hamiltonian \eqref{eq:Ham} and $n_F(\epsilon)=1/(e^{\epsilon/T}+1)$ is the Fermi function. For numerical calculations, we use the following forms of the susceptibilities if the energy denominators vanish (or are close to vanishing)
\begin{gather}
\dot{\pi}^{ph}(E,E+\delta)\simeq\frac{1}{T^2}\frac{1}{4\cosh^2\frac{x}{2}}(1-x\tanh\frac{x}{2})\\
\dot{\pi}^{pp}(E,-E+\delta)\simeq \frac{1}{T^2}\frac{1}{4\cosh^2\frac{x}{2}}(-1+x\tanh\frac{x}{2}),
\end{gather} where $x=E/T$. As is commonly done \cite{Platt2013review}, we have neglected the frequency-dependence of the vertex, the self-energy correction and the contribution from higher order vertices (six-point vertex and beyond) in the FRG equations. For weak enough interactions, these will be good approximations. 

Consider the following form of the FRG equations in terms of the full antisymmetric vertex function $\gamma$
\begin{align}
    \dot{\gamma}(1,2,3,4)=\frac{1}{2}\sum_k\bigg[-\dot{\pi}_{pp}(k,k')&\gamma(1,2,k,k')\gamma^*(3,4,k,k')\\
    +2\dot{\pi}_{ph}(k,k')&\gamma^*(3,k,1,k')\gamma(2,k,4,k')\\
    -2\dot{\pi}_{ph}(k,k')&\gamma^*(3,k,2,k')\gamma(1,k,4,k')\bigg]
\end{align}
where band/flavour indices are implicit, and the momentum $k'$ is fixed by momentum conservation in each term. The derivative susceptibilities $\dot{\pi}$ have been redefined here so that its momentum arguments are the ones that directly enter the individual propagators. In the presence of $U(1)_S$ spin symmetry (i.e.~for a Zeeman field, or Ising SOC), we can decompose the full vertex into the following independent spin components
\begin{gather}
    U=\gamma^{\uparrow\uparrow\uparrow\uparrow}, \quad D=\gamma^{\downarrow\downarrow\downarrow\downarrow},\quad
    C=\gamma^{\uparrow\downarrow\uparrow\downarrow}\\
    \pi^U=\pi^{\uparrow\uparrow},\quad \pi^D=\pi^{\downarrow\downarrow},\quad \pi^C=\pi^{\uparrow\downarrow}
\end{gather}
where we have also introduced convenient notation for the susceptibilities. Note that e.g. $C$ does not enjoy the full index exchange properties of $\gamma$. In terms of these, the FRG equations become
\begin{align}
    \dot{U}(1,2,3,4)=\frac{1}{2}\sum_k\bigg[-\dot{\pi}_{pp}^U(k,k')&U(1,2,k,k')U^*(3,4,k,k')\\
    +2\dot{\pi}^U_{ph}(k,k')&U^*(3,k,1,k')U(2,k,4,k')\\
    +2\dot{\pi}^D_{ph}(k,k')&C^*(3,k,1,k')C(2,k,4,k')\\
    -2\dot{\pi}^U_{ph}(k,k')&U^*(3,k,2,k')U(1,k,4,k')\\
    -2\dot{\pi}^D_{ph}(k,k')&C^*(3,k,2,k')C(1,k,4,k')\bigg]
\end{align}
\begin{align}
    \dot{D}(1,2,3,4)=\frac{1}{2}\sum_k\bigg[-\dot{\pi}_{pp}^D(k,k')&D(1,2,k,k')D^*(3,4,k,k')\\
    +2\dot{\pi}^D_{ph}(k,k')&D^*(3,k,1,k')D(2,k,4,k')\\
    +2\dot{\pi}^U_{ph}(k,k')&C^*(k,3,k',1)C(k,2,k',4)\\
    -2\dot{\pi}^D_{ph}(k,k')&D^*(3,k,2,k')D(1,k,4,k')\\
    -2\dot{\pi}^U_{ph}(k,k')&C^*(k,3,k',2)C(k,1,k',4)\bigg]
\end{align}
\begin{align}
    \dot{C}(1,2,3,4)=\frac{1}{2}\sum_k\bigg[-\dot{\pi}_{pp}^C(k,k')&C(1,2,k,k')C^*(3,4,k,k')\\
    -\dot{\pi}_{pp}^{-C}(k,k')&C(1,2,k',k)C^*(3,4,k',k)\\
    +2\dot{\pi}^U_{ph}(k,k')&U^*(3,k,1,k')C(k,2,k',4)\\
    +2\dot{\pi}^D_{ph}(k,k')&C^*(3,k,1,k')D(2,k,4,k')\\
    -2\dot{\pi}^{-C}_{ph}(k,k')&C^*(3,k,k',2)C(1,k,k',4)\bigg]
\end{align}
where $\pi^{-C}=\pi^{\downarrow\uparrow}$. We label the terms U1, U2, etc in order. Then there are several simplifying relations. The pairs (U2,U4), (U3,U5), (D2,D4), (D3,D5) are negative transposes (swap momentum/band indices 1 and 2) of each other. C1 and C2 are identical. So we only need to explicitly do 10 summations.

\section{Momentum Patching}

In this section, we discuss some considerations regarding momentum patching. In the most naive implementation of FRG, the momenta appearing as arguments of the susceptibilities $\pi_{pp}$ and $\pi_{ph}$ as well as the momentum arguments of the four-point vertex all lie on a grid covering the BZ. However, with such a discretization of the FRG equations, it is not possible to get a sufficient resolution of the Fermi surface.  Typically the FRG equation is not solved for all possible momenta $\bp_1,\bp_2,\bp_3$. Instead some coarse-graining procedure is performed, where clumps of momenta are grouped into patches.  This leads to a refined method often referred to as patch FRG. In this implementation, the momenta appearing inside the susceptibilities are evaluated on a very fine mesh, which is necessary in order to resolve the sharp features of the susceptibilities appearing at low temperatures. On the other hand the momenta appearing as arguments of the four-point vertex are picked to lie exactly on the Fermi surface. These momenta are expected to be the only momenta relevant at the lowest energy scales, since the components of the four-point vertex on the Fermi surface is what enters the gap equation.

Let us set up some conventions for the patches. We imagine that we have a collection of Bloch momenta $\mathcal{D}_{\text{BZ}}$ that is common to all bands. Note that these Bloch momenta only cover a region of the BZ close to the $K$-points. For each band $a$, we divide $\mathcal{D}_{\text{BZ}}$ into $N_p$ non-overlapping patches indexed by $m$. So the pair of indices $(a,m)$ labels a patch, defined as the particular collection of momenta $\mathcal{D}_{(a,m)}$ (so patches from different bands can overlap). We say that $\bk$ belongs to patch $(a,m)$ if $\bk\in\mathcal{D}_{(a,m)}$. Furthermore, each patch is associated with a representative momentum $\bm{K}_{(a,m)}$ called the patch momentum. Typically the patch momentum is close to the FS and is near the centre of the patch (in angular terms). Sums over band and momentum can be decomposed as follows
\begin{equation}
    \sum_{\bp,a}=\sum_{a}\sum_{m}\sum_{\bk\in\mathcal{D}_{(a,m)}}.
\end{equation}

Given some momentum patching, we can consider coarse-graining the coupling functions so that they depend on the patch indices rather than momenta. The flow equation \eqref{eq:FRG_flow_app} becomes
\begin{align}
\label{eq:FRG_flow_p}
    \dot\gamma_{ab;cd}(m_1,m_2,m_3)=\sum_{{x,y}}\sum_{m=1}^{N_p}&\bigg[-\dot\Pi^{pp}_{xy}(m,\bm{K}_{(a,m_1)}+\bm{K}_{(b,m_2)})
    \gamma_{ab;xy}(m_1,m_2,m)\gamma^{ *}_{cd;xy}(m_3,m_4,m)\nonumber
   \\&+2\dot\Pi^{ph}_{xy}(m,\bm{K}_{(a,m_1)}-\bm{K}_{(c,m_3)})\gamma^{*}_{cx;ay}(m_3,m,m_1)\gamma_{bx;dy}(m_2,m,m_4)
   \\&-2\dot\Pi^{ph}_{xy}(m,\bm{K}_{(b,m_2)}-\bm{K}_{(c,m_3)})\gamma^{*}_{cx;by}(m_3,m,m_2)\gamma_{ax;dy}(m_1,m,m_4)\bigg]\nonumber,
\end{align}
where $m_4$ is uniquely determined by all of the other indices. The other momenta in the coupling function are fixed to the patch momenta, thereby determining $\bp_4$, which lands in one of the patches, by momentum conservation. This momentum is then projected to the corresponding patch momentum. Above, we have defined partially integrated susceptibilities
\begin{gather}
    \Pi_{xy}(m,\bq)=\frac{1}{A}\sum_{\bk\in\mathcal{D}_{(x,m)}}\pi_{xy}(\bk,\bq).
\end{gather}
To construct the patching, we pick $N_p$ equally spaced points along the Fermi surface as the representative patch momenta. The patches are then obtained by the Voronoi tessellation of these patch momenta. 

In order to make the problem computationally tractable, we must exploit the symmetries in the problem. We have already exploited the spin-$U_S(1)$ symmetry, as outlined in the re-writing of the FRG equations above. In principle, one could also exploit the valley-$U_V(1)$ symmetry and thereby achieve a further speedup, however, this turns out not to be necessary for the system size we study. Let us first focus on the case with a magnetic field. Then the other symmetries we use are time-reversal $\mathcal{T}$ and $C_3$. We pick a patching that is consistent with these symmetries, in particular we pick
\begin{align}
    \bm{K}_{(a,(m+N_p/3)\mathrm{mod}N_p)}&=C_3\bm{K}_{(a,m)}\\
    \bm{K}_{(\tau=+,\sigma,m)}&=-\bm{K}_{(\tau=-,\sigma,m)}.
\end{align}
The four-point vertex then satisfies
\begin{align}
    \gamma_{ab;cd}(m_1+\frac{N_p}{3},m_2,m_3)&=\gamma_{ab;cd}(m_1,m_2-\frac{N_p}{3},m_3-\frac{N_p}{3})\\
    \gamma_{\sigma_a\tau_a\sigma_b\tau_b;\sigma_c\tau_c\sigma_d\tau_d}(m_1,m_2,m_3)&=\gamma_{\sigma_a\bar\tau_a\sigma_b\bar\tau_b;\sigma_c\bar\tau_c\sigma_d\bar\tau_d}(m_1,m_2,m_3)^*,
\end{align}
where all the patch indices are understood modulo $N_p$ and $\bar\tau$ denotes the valley opposite to $\tau$. After imposing the $U_S(1)$, $\mathcal{T}$ and $C_3$ symmetries, the four-point vertex which initially contained $2^8N_p^3$ components is reduced to $2^3N_p^3$ independent components.

\subsection{Gap equation}
We consider particle-particle orders and include explicitly momentum and valley labels ($\tau=\pm$). Consider the interacting Hamiltonian
\begin{equation}
    H=\sum_{k\tau ab}h_{ab}(k,\tau)c^\dagger_{k\tau a}c_{k\tau b}+\frac{1}{2A}\sum_{\{a\}\tau\tau'\{k\}}V^{\tau\tau'}_{abcd}(k_1,k_2,k_3)c^\dagger_{k_1\tau a}c^\dagger_{k_2\tau' b}c_{k_4\tau' d}c_{k_3\tau c}
\end{equation}
where spins/bands are lumped into the index $a$. This resembles a density-density interaction in valley space, but intervalley scattering terms can be folded into the above (since we have not imposed any spin structure). We consider intervalley pairing and define the anomalous density
\begin{equation}
    \kappa_{ab}(k)\equiv\langle c_{-k,-,b}c_{k,+,a}\rangle,\quad \kappa_{ab}^*(k)\equiv\langle c^\dagger_{k,+,a}c^\dagger_{-k,-,b}\rangle.
\end{equation}
We rewrite the interaction
\begin{align}
    H_{\text{int}}&=\frac{1}{A}\sum_{\{a\}\{k\}}\left[\frac{1}{2}V^{+-}_{abcd}(k_1,k_2,k_3)+\frac{1}{2}V^{-+}_{badc}(k_2,k_1,k_4)\right]c^\dagger_{k_1+a}c^\dagger_{k_2-b}c_{k_4-d}c_{k_3+c}+\ldots\\
    &\equiv \frac{1}{A}\sum_{\{a\}\{k\}}\Gamma_{abcd}(k_1,k_2,k_3)c^\dagger_{k_1+a}c^\dagger_{k_2-b}c_{k_4-d}c_{k_3+c}+\ldots
\end{align}
where the dots represent intravalley terms which are not involved in pairing, and we have the relation $\Gamma_{abcd}(k_1,k_2,k_3)=\Gamma_{cdab}^*(k_3,k_4,k_1)$. Now decouple $c_{-k,-,a}c_{k,+,b}=\kappa_{ba}(k)+\left[c_{-k,-,a}c_{k,+,b}-\kappa_{ba}(k)\right]$ and neglect quadratic fluctuations
\begin{align}
    H_{\text{int}}\rightarrow& -\frac{1}{A}\sum_{\{a\}kk'}\kappa_{ab}^*(k)\Gamma_{abcd}(k,-k,k')\kappa_{cd}(k')\\
    &+\frac{1}{A}\sum_{\{a\}kk'} c^\dagger_{k,+,a}c^\dagger_{-k,-,b}\Gamma_{abcd}(k,-k,k')\kappa_{cd}(k')\\
    &+\frac{1}{A}\sum_{\{a\}kk'}c_{-k,-,b}c_{k,+,a}\Gamma_{abcd}^*(k,-k,k')\kappa^*_{cd}(k')
\end{align}
Define the BCS scattering vertex $U$ and the gap function $\Delta$
\begin{gather}
    U_{abcd}(k,k')=\Gamma_{abcd}(k,-k,k')=U_{cdab}^*(k',k)\\
    \Delta_{ab}(k)=\frac{1}{A}\sum_{cdk'}U_{abcd}(k,k')\kappa_{cd}(k')
\end{gather}
i.e. $\bm{\Delta}=\frac{1}{A}U\cdot\bm{\kappa}$ if we treat $(kab),(k'cd)$ as matrix indices. The interaction part of the mean-field BCS Hamiltonian becomes
\begin{equation}
    H^{\text{int,BCS}}=\sum_{kab}c^\dagger_{k,+,a}c^\dagger_{-k,-,b}\Delta_{ab}(k)+\sum_{kab}c_{-k,-,b}c_{k,+,a}\Delta^*_{ab}(k)-A\bm{\Delta}^*\cdot U^{-1}\cdot \bm{\Delta}.
\end{equation}
Define the Nambu spinor $\psi_a(k)=[c_{k,+,a},c_{-k,-,a}]^T$, which leads to the total mean-field Hamiltonian
\begin{equation}
    H^{\text{MF}}=\sum_k\bm{\psi}^\dagger(k)\begin{pmatrix}
    h(k,+) & \Delta(k) \\ \Delta^\dagger(k) & -h(-k,-)
    \end{pmatrix}\bm{\psi}(k)-A\bm{\Delta}^*\cdot U^{-1}\cdot \bm{\Delta}
\end{equation}
where the vector notation in $\bm{\psi}$ is in Nambu and band space. The matrix above is denoted $h^{\text{BCS}}$. In the path-integral, we have the action
\begin{equation}
    S=\int_0^\beta d\tau \left[\sum_k\bar{\bm{\psi}}(\partial_\tau+h^{\text{BCS}})\bm{\psi}(k) -A\bar{\bm{\Delta}}\cdot U^{-1}\cdot \bm{\Delta} \right].
\end{equation}
We integrate out the fermion fields, assuming a static pairing field, leading to the effective free energy
\begin{equation}
    \mathcal{F}(\Delta,\bar{\Delta})=-T\sum_{k,n}\text{tr}\ln \left[-i\omega_n+h^{\text{BCS}}(k)\right]-A\bar{\bm{\Delta}}\cdot U^{-1}\cdot \bm{\Delta}.
\end{equation}
The trace above acts in Nambu and band (including spin) space. We next minimize this with respect to $\bar{\Delta}_{ab}(p)$. 

Consider first the $\text{tr}\ln$ term. We need two facts: the identity $\frac{\delta}{\delta \phi}\text{tr}\ln M^{-1}=\text{tr}\left(M\frac{\delta}{\delta\phi}M^{-1}\right)$ and the fact that the top-right component of the block matrix $\begin{pmatrix}A & B\\C & D\end{pmatrix}^{-1}$ is $-A^{-1}B(D-CA^{-1}B)^{-1}$. Applying these, we obtain
\begin{align}
    \frac{\delta \text{tr}\ln}{\delta\bar{\Delta}_{ab}}&=
    T\sum_n\left[\left[-i\omega_n+h(p,+)\right]^{-1}\Delta(p)\left[-i\omega_n-h(-p,-)-\bar{\Delta}(p)\frac{1}{-i\omega_n+h(p,+)}\Delta(p)\right]^{-1}\right]_{ab}\\
    &\simeq T\sum_n\left[\frac{1}{-i\omega_n+h(p,+)}\Delta(p)\frac{1}{-i\omega_n-h(-p,-)}\right]_{ab}
\end{align}
where we have linearized in the second line. If our basis corresponds to a band basis where $h$ is diagonal, we obtain
\begin{equation}
    \frac{\delta \text{tr}\ln}{\delta\bar{\Delta}_{ab}}=-\pi^{\text{pp},+-}_{ab}(p,0)\Delta_{ab}(p)
\end{equation}
where $\pi^{\text{pp},\tau\tau'}_{ab}(k,q)=T\sum_n G_{\tau a}(k,\omega_n)G_{\tau',b}(-k+q,-\omega_n)$. 

The second term in the free energy has variation $-A\sum_{kcd}[U^{-1}]_{abcd}(p,k)\Delta_{cd}(k)$ leading to the gap equation
\begin{equation}
    \Delta_{ab}(p)=-\frac{1}{A}\sum_{cd,k}U_{abcd}(p,k)\pi^{\text{pp},+-}_{cd}(k,0)\Delta_{cd}(k).
\end{equation}

\end{appendix}
\end{document}